\documentclass[12pt]{article}
\usepackage{amsmath}
\usepackage{graphics}
\usepackage{latexsym}

\usepackage [ansinew]{inputenc}
\usepackage {hyperref}
\begin{document}

\vskip 2cm
\title{Improved Stability for Pulsating Multi-Spin String Solitons}
\author{\\
A. Khan${}^{|}$ and A.L. Larsen${}^{*}$}
\maketitle
\noindent
{\em Physics Department, University of Southern Denmark,
Campusvej 55, 5230 Odense M,
Denmark}
\begin{abstract}
\hspace*{-6mm}We
derive and analyse, analytically and numerically,
the equations for perturbations around the pulsating
two-spin string soliton in $AdS_5\times S^5$.
We show that the pulsation in $S^5$
indeed improves the stability properties of the two-spin string soliton in
$AdS_5$,
in the sense that the more pulsation we have,
the higher spin we can allow and still have stability.
\end{abstract}

\vskip 6cm
\noindent
$^{|}$Electronic address: patricio@fysik.sdu.dk\\
$^{*}$Electronic address: all@fysik.sdu.dk
\baselineskip=1.5em

\newpage
\section{Introduction}
\setcounter{equation}{0}
The result of semi-classical quantization of long spinning strings in Anti de 
Sitter space is that the energy $E$ scales with
the spin $S$, $E\sim S$ (for large $S$).
This
result, which is independent of the dimensionality of
Anti de Sitter space, was
originally obtained a decade ago by considering fluctuations around
the string center
of mass
\cite{de vega}. It holds even classically, as was immediatly shown
by considering rigidly rotating strings \cite{inigo}, and is due to the 
constant
curvature of Anti de Sitter space.
These findings have recently received a lot of attention in connection with 
the
conjectured
duality \cite{maldacena,gubser,witten} between super string theory on ${\rm
AdS_5}\times {S^5}$ and
${\cal N}=4$ SU(N)
super Yang-Mills theory in Minkowski space. In the case of rigidly	 rotating
strings, it was
noticed \cite{klebanov} that the
subleading term is logarithmic in the spin $S$
\begin{equation}
E-S\sim \ln(S)\ \end{equation}
which is essentially the same behavior as found for certain operators on
the gauge theory side
\cite{gross,georgi,floratos, korchemsky, dolan}.
The $E$-$S$ relationship has been further investigated in a number of
papers including \cite{tseytlin,tseytlin2,russo,armoni,
mandal,minahan,
barbon,axenides,alishabiba,buchel,rashkov,ryang,bozhilov,bozhilov2,patricio,smed,col,col1}
for various string configurations.

More recently it was discovered that multi spin solutions could also be
constructed easily \cite{tseytlin3}. Among others, a relatively
simple solution was found in $AdS_5$ describing a string which is located
at a point in the radial direction, winding around an
angular direction and spinning with equal angular momentum in two
independent planes. In the long string limit it was shown that
\begin{equation}
E-2S\sim S^{1/3}
\label{eq1.2}
\end{equation}
Multi spin solitons in $AdS_5\times S^5$ have been
further investigated in a number of papers including
\cite{engquist,beisert,tseytlin4,park,tseytlin5,tseytlin6,
tseytlin7,tseytlin8,stefan,kim,tseytlin9,patricio1,mohsen,dimov2,stefanski,engquist2,
kato,ryang2,belucci,ryang3,ryshov,mosaffa}. However, most of these papers have 
dealt with multi spin solutions in the $S^5$ part of $AdS_5\times S^5$.

The main problem with all these multi spin solutions, is that they tend to be 
classically unstable for large spin. For instance, the simple two-spin 
solution
leading to (\ref{eq1.2}) is only stable for \cite{tseytlin3}

\begin{eqnarray}
S\leq\frac{5\sqrt{7}}{8\sqrt{2}H^2\alpha'}
\end{eqnarray}
This means, strictly speaking, that the result (\ref{eq1.2}) can not be 
trusted,
since it was derived under the assumption that $S>>(H^2\alpha')^{-1}$.

The purpose of the present paper is to show that pulsation improves the 
stability
properties for multi spin string solitons. More precisely, we take the simple
two-spin string solution in $AdS_5$ \cite{tseytlin3}, couple it with pulsation 
in $S^5$ and consider small perturbations around it. We show that this 
solution
has better stability properties than the non-pulsating one.

The paper is organised as follows. In section 2, we give a short review of the 
pulsating two-spin solution in $AdS_5\times S^5$ \cite{patricio}. Section 3 is 
devoted to a general
discussion of perturbations, and in section 4 we derive the equations of 
motion for
the physical perturbations around the pulsating two-spin solution. In section 
5, we solve the equations analytically for a few tractable cases, while 
leaving numerical analysis of the general case for section 6. Finally, in 
section 7, we present our conclusions.

\section{ Pulsating Multi Spin Solutions}
\setcounter{equation}{0} In this section, we set our conventions and 
notations, and review the
pulsating
two-spin string soliton \cite{patricio}.

We have altogether for
$AdS_5\times S^5$, the line-element
\begin{eqnarray}
ds^2&=&-(1+H^2r^2)dt^2+\frac{dr^2}{1+H^2r^2}+r^2(d\beta^2+\sin^2\beta
d\phi^2+\cos^2\beta d\tilde{\phi}^2 )
\nonumber \\
&+& \frac{1}{H^2}(d\theta^2+\sin^2\theta d\psi^2+
\cos^2 \theta (d\psi_1^2+\sin^2\psi_1 d\psi_2^2+\cos^2\psi_1 d\psi_3^2 ))
\end{eqnarray}
The 't
Hooft coupling in this notation is
$\lambda =(H^2\alpha')^{-2}$, where $(2\pi\alpha')^{-1 }$ is the string
tension.
The radius of $S^5$ is $H^{-1}$.

In the standard parametrisation of $S^3$ we have the ranges
$\beta\in[0,\pi/2]$, $\phi\in[0,2\pi]$, $\tilde{\phi}\in[0,2\pi]$,
but here we shall use the alternative ranges
$\beta\in[0,2\pi]$, $\phi\in[0,\pi]$, $\tilde{\phi}\in[0,\pi]$.

We take the Polyakov action
\begin{equation} \label{23}
S=-\frac{1}{4\pi\alpha'}\int d\tau d\sigma \sqrt{-h}
h^{\alpha\beta} G_{\mu\nu}
X^\mu_{,\alpha}X^\nu_{,\beta}
\end{equation}
using the
conformal gauge
\begin{equation} \label{22}
G_{\mu\nu}\dot X^{\mu}{X'}^{\nu}=0,\ \ \ G_{\mu\nu}
(\dot X^\mu\dot X^\nu+{X'}^\mu {X'}^\nu)=0
\end{equation}
where dot and prime denote derivatives with respect to the world-sheet
coordinates
$\tau$ and $\sigma$, such that
the equations of motion are
\begin{equation} \label{26}
\ddot X^{\mu}-{X''}^\mu + \Gamma^\mu_{\rho\sigma}(\dot
X^\rho\dot X^\sigma -{X'}^\rho {X'}^\sigma)= 0
\end{equation}
The ansatz for the pulsating two-spin string soliton
is
\begin{eqnarray}
&&t=c_0\tau,\ \ \ r=r_0, \ \ \ \beta =\sigma ,\ \ \
\phi=\omega\tau, \ \ \
\tilde{\phi}=\omega\tau,\nonumber \\
&& \theta=\theta(\tau), \ \ \
\psi=\sigma,\ \ \
\psi_1=
\psi_{10}
,\ \ \ \psi_2=\psi_{20},\ \ \ \psi_3=\psi_{30}
\end{eqnarray}
where $(c_0,r_0,\omega,\psi_{10},
\psi_{20},
\psi_{30}
)$ are arbitrary constants.
This is a circular
string in $AdS_5$, spinning in two different
directions. It is also a circle in $S^5$, but pulsating there.
The $r$ and $\theta $ equations become
\begin{equation}
\omega^2=1+H^2c_0^2
\end{equation}
\begin{equation}
\label{2.7}
\ddot\theta+\sin\theta\cos\theta=0
\end{equation}
while the non-trivial conformal gauge constraint is
\begin{equation}
\dot\theta^2+\sin^2\theta-H^2(c_0^2-2r_0^2)=0
\end{equation}
The $\theta $ equation and constraint are solved by
\begin{eqnarray}\label{2.9}
\sin\theta(\tau)&=& \left\{ \begin{array}{ll}
A{\mbox {sn}}(\tau |A^2 )
\ , \ \ &A\leq 1
\\
{\mbox {sn}}(A\tau |1/A^2)\ ,\ \ & A\geq 1 \end{array} \right.
\end{eqnarray}
where
\begin{equation} \label{2.10}
A=H\sqrt{c_0^2-2r_0^2}
\end{equation}
Notice that (\ref{2.7}) is equivalent to the equation for a simple plane
pendulum. It means that we have two types of motion, and a limiting case.
More precisely, for $A<1$ the string oscillates on one hemisphere, for $A=1$
it starts at one of the poles and approaches the
equator for $\tau\rightarrow\infty$,
while for $A>1$ it oscillates between the poles.

The energy $E$ and
the 2 spins
$S_1=S_2\equiv S$ are easily computed
\begin{eqnarray}\label{7}
E&=&\frac{(1+H^2r_0^2)\sqrt{2H^2r_0^2+A^2}}{H\alpha'}\\
S&=&\frac{r_0^2}{2\alpha'}\sqrt{2r_0^2H^2+A^2+1}\label{2.12}
\end{eqnarray}
For short strings (say $Hr_0<<1$) we get
\begin{eqnarray}\label{7}
S&\approx &\frac{r_0^2}{2\alpha'}\sqrt{A^2+1} \left(1+\frac{H^2r_0^2}{A^2+1}
\right)
\end{eqnarray}
such that
\begin{eqnarray}
H^2r_0^2\approx \frac{2H^2\alpha' S}{\sqrt{A^2+1}}-
\frac{4H^4{\alpha'}^2 S^ 2}{(A^2+1)^ 2}
\end{eqnarray}
which inserted into $E$ gives
\begin{equation}
E(S,A)\approx \frac{1}{H\alpha'}\left(1+\frac{2H^2\alpha'S}{\sqrt{A^2+1}}-
\frac{4H^4{\alpha'}
^2S^2}{(A^2+1)^2}\right)\sqrt{\frac{4H^2\alpha'S}{\sqrt{A^2+1}}-
\frac{8H^4{\alpha'} ^2S^2}{(A^2+1)^2}+A^2}
\end{equation}
For $A=0$ we get
\begin{eqnarray}
E\approx \frac{2\sqrt{S}}{\sqrt{\alpha'}}(1+H^2\alpha'S)
\end{eqnarray}
which to leading order is just the Minkowski result $\alpha' E^2=2(2S)$.
For $A=1$ we get
\begin{eqnarray}
E\approx \frac{1}{H\alpha'}(1+2^{3/2}H^2\alpha'S)
\end{eqnarray}
and for $A>>1$
\begin{eqnarray}
E\approx \frac{1}{H\alpha'}(A+2H^2\alpha'S)
\end{eqnarray}
For long strings (say $Hr_0>>1$) we get
\begin{eqnarray}
E/H-2S
&\approx &\frac{\sqrt{2H^2r_0^2+A^2}}{H^2\alpha'} -
\frac{H^2r_0^2}{2H^2\alpha' \sqrt{2H^2r_0^2+A^2}}
\label{4.15}
\end{eqnarray}
Now we have to distinguish between different cases. If $Hr_0>>A$, we get
from (\ref{2.12})
\begin{eqnarray}
S&\approx
&\frac{Hr_0^3}{\sqrt{2}\alpha'}\left(1+\frac{A^2+1}{4H^2r_0^2}\right)
\end{eqnarray}
such that
\begin{eqnarray}
Hr_0\approx (\sqrt{2}H^2\alpha' S)^{1/3}-\frac{A^2+1}{12(\sqrt{2}H^2\alpha'
S)^{1/3} }
\end{eqnarray}
which, when inserted into (\ref{4.15}), gives
\begin{eqnarray}
E/H-2 S\approx \frac{3(\sqrt{2}H^2\alpha'S)^{1/3}}{2^{3/2}H^2\alpha'}+
\frac{4A^2-1}{2^{7/2}H^2\alpha'(\sqrt{2}H^2\alpha'S)^{1/3}}
\end{eqnarray}
This result is valid for $H^2\alpha' S>>\{1, \ A^3\}$, and therefore holds
in particular for $A=0$ and $A=1$.
Notice also that the pulsation only gives a contribution to the non-leading
terms, in this limit.
On the other hand, if $Hr_0<<A$ we get from eq.(\ref{2.12})
\begin{eqnarray}\label{7}
S&\approx &\frac{Ar_0^2}{2\alpha'}\left(1+ \frac{r_0^2H^2}{A^2}\right)
\end{eqnarray}
such that
\begin{eqnarray}\label{7}
H^2r_0^2\approx \frac{2H^2S\alpha'}{A}-\frac{4S^2{\alpha'}^2H^4}{A^4}
\end{eqnarray}
and insertion into (\ref{4.15}) gives
\begin{eqnarray}\label{7}
E/H-2 S\approx \frac{A}{H^2\alpha'}+\frac{S}{A^2}
\end{eqnarray}
which holds for $1<<H^2\alpha' S<<A^3$. Thus, in this limit, the pulsation
completely changes the
scaling relation.

For $A=0$ all the results of this section reduce to those obtained in
\cite{tseytlin3}, but are
otherwise quite
different. The main problem is that the string soliton is
generally
not stable. For $A=0$ it was shown in \cite{tseytlin3} that there is only
stability for
\begin{eqnarray}
S\leq\frac{5\sqrt{7}}{8\sqrt{2}H^2\alpha'}
\end{eqnarray}
The purpose of the following sections is to find the condition for
stability for arbitrary $A$.

\section{Perturbations}
\setcounter{equation}{0}
There are many ways to discuss the perturbations around string solitons.
One way is to use the Nambu-Goto action

\begin{equation}
S=-\frac{1}{2\pi\alpha'}\int d\tau d\sigma \sqrt{-det(
G_{\mu\nu}
X^\mu_{,\alpha}X^\nu_{,\beta})}
\end{equation}
and make variations $\delta X^\mu$ to obtain $\delta^2 S$,
immediately fixing two of them so as to take care of the gauge invariance.
This is the most often used method in recent papers. The disadvantages are
that it is not world-sheet covariant,
and that the kinetic energy terms for the perturbations usually come out in
a very complicated form.

Another way is to take the Polyakov action in conformal gauge; i.e., one
makes
a variation of (\ref{22})
and two variations of (see \cite{san,san1,san2})
\begin{equation}
S=\frac{1}{4\pi\alpha'}\int d\tau d\sigma G_{\mu\nu}
(\dot X^\mu \dot X^\nu - {X'}^\mu {X'}^\nu)
\end{equation}
This approach is also used in many
papers, but the problem here is that one can not usually solve the variation 
of
(\ref{22}),
making it is very difficult to separate the physical and unphysical
perturbations.

Here we will use the Polyakov action (\ref{23}); i.e.,
make two variations $\delta X^\mu$ and $\delta h_{\alpha\beta}$.
To ensure that we have purely physical perturbations, we shall consider only 
perturbations which are normal to the string
world-sheet. This approach is
in general world-sheet
covariant, and we will eventually get the kinetic
energy terms in a very simple form (see
\cite{frolov,guven,guven2,kar2}).

First introduce 8 normal vectors to the world-sheet ($i=1, 2,...,8$)
fulfilling
\begin{equation} \label{430}
G_{\mu\nu}N_i^{\mu}N_j^{\nu}=\delta _{ij} ,\;\;\;\;
G_{\mu\nu}N_i^{\mu}X^{\nu}_{,\alpha}=0
\end{equation}
as well as the completeness relation
\begin{equation} \label{431}
G^{\mu\nu}=\frac{ 2h^{\alpha\beta}}{h^{\gamma\delta}g_{\gamma\delta}}
X^\mu_{,\alpha}X^{\nu}_{,\beta} +\delta ^{ij}N^\mu_i N^{\nu}_j
\end{equation}
where $g_{\alpha\beta}$ is the induced metric on the world-sheet.
Then define the second fundamental form and normal fundamental form
\begin{equation}
K^i_{\alpha\beta}=N^i_{\mu}X^{\rho}_{,\beta}\nabla_{\rho}
X^\mu_{,\alpha},\ \ \ \
\mu^{ij}_{\alpha}=N^i_{\mu}X^{\rho}_{,\alpha}\nabla_{\rho}N^{j\mu}
\end{equation}
where $\nabla_\rho$ is the covariant derivative with respect to the
metric $G_{\mu\nu}$.

Since $(X^\mu_{,\alpha},\
N^\mu_i)$ is a
basis for spacetime (at least
locally), we can write
\begin{equation} \label{421}
\delta X^\mu=N^\mu_i\phi^i+X^\mu_{,\alpha}\psi^{\alpha}
\end{equation}
The $\psi^\alpha$ are just reparametrisations, and it can then be shown
that the $\phi^i$ fulfil the equations
(see \cite{frolov,guven,guven2,kar2})

\begin{eqnarray} \label{440}
\left(\delta^{kl}h^{\alpha\beta}D_{ik\alpha}D_{lj\beta}
+
\frac{2}{h^{\epsilon\zeta}g_{\epsilon\zeta}}K^{\alpha\beta}_i
K_{j\alpha\beta}
+h^{\alpha\beta}R_{\mu\rho\nu\sigma}N^\mu_iN^{\nu}_jX^{\rho}_{,\alpha}X^{\sigma
}_{,\beta}\right)
\phi^j=0
\end{eqnarray}
where we introduced $D_{ij\alpha}=\delta
_{ij}D_{\alpha}+\mu_{ij\alpha}$, and $D_\alpha$ is the covariant derivative
with respect to the Polyakov metric.

\section{Stability of Pulsating Two-Spin String Soliton}
\setcounter{equation}{0}

As the unperturbed string we take the pulsating two-spin solution in
$AdS_5\times S^5$
from section 2.
That is, $h_{\alpha \beta }=\eta_{ \alpha \beta }=diag(-1,\ 1)$ as well as
\begin{equation}
{X}^\mu=(c_0\tau,r_0,\sigma ,\omega\tau,\omega\tau,\theta(\tau
),\sigma,\psi_{10},\psi_{20},\psi_{30})
\end{equation}
The tangent vectors are
\begin{equation}
\dot{X}^\mu=(c_0,0,0,\omega,\omega ,\dot{\theta},0,0,0,0)
\end{equation}
\begin{equation}
X'^\mu=(0,0,1,0,0,0,1,0,0,0)
\end{equation}
The induced metric on the world-sheet is
\begin{eqnarray}
g_{\tau\sigma}=0,\ \ \ \
g_{\tau\tau}=- g_{\sigma\sigma}
=
-H^{-2}(H^2r_0^2+\sin^2\theta )
\end{eqnarray}
There are 8 normal vectors $N^i_\mu$
fulfilling eqs.(\ref{430})-(\ref{431}). They can be chosen as
\begin{eqnarray}
N^1_\mu&=& \frac{ r_0\sqrt{1+H^2r_0^2}}{\sqrt{c_0^2-r_0^2}}
(\omega,0,0,-c_0\sin^2\sigma ,-c_0\cos^2\sigma ,0,0,0,0,0) \\
N^2_\mu&=& r_0\sin\sigma\cos\sigma (0,0,0,1,-1,0,0,0,0,0) \\
N^3_\mu& =& (1+H^2r_0^2)^{-1/2}(0,1,0,0,0,0,0,0,0,0)\\
N^4_\mu&=&\frac{r_0\sin\theta}{\sqrt{H^2r_0^2+\sin^2\theta}}
(0,0,1,0,0,0,-1,0,0,0) \\
N^5_\mu&=& \frac{1}{\sqrt{c_0^2-r_0^2}\sqrt{H^2r_0^2 +\sin^2\theta}}\\
&&\left( {-c_0}(1+H^2r_0^2)\dot{\theta},0,0,\omega r_0^2
\sin^2\sigma\dot{\theta},
\omega r_0^2 \cos^2\sigma\dot{\theta},
c_0^2-r_0^2 ,0,0,0,0\right)\nonumber\\
N^6_\mu&=&H^{-1 }\cos \theta (0,0,0,0,0,0,0,1,0,0) \\
N^7_\mu&=& H^{-1 }\cos\theta\sin\psi_{10}(0,0,0,0,0,0,0,0,1,0) \\
N^8_\mu&=& H^{-1 }\cos\theta\cos\psi_{10} (0,0,0,0,0,0,0,0,0,1)
\end{eqnarray}
Notice that the first 3 only have components in $AdS_5$, while the last
3 only have components in $S^5$. It is then straightforward to compute the
second fundamental form,
the normal fundamental form and the projections of the Riemann tensor,
appearing in (\ref{440}).
For convenience, they are listed in the appendix. We now have all
the ingredients to write down the equations (\ref{440}) for the
perturbations

\begin{eqnarray}\label{4.13}
&&-\ddot \phi^1
+{\phi''}^1+\frac{2c_0\omega}{\sqrt{c_0^2-r_0^2}}\ \dot\phi^3
+\frac{2c_0\sqrt{1+H^2r_0^2}}{\sqrt{c_0^2-r_0^2}}\
\phi'^2=0\\
&&-\ddot\phi^2 +{\phi''}^2-\frac{2\omega\sin\theta}{
\sqrt{H^2r_0^2+\sin^2\theta}}\dot\phi^4
-\frac{2c_0\sqrt{1+H^2r_0^2}}{\sqrt{c_0^2-r_0^2}}\phi'^1\nonumber\\
&+&
\frac{2r_0\omega\dot\theta}{\sqrt{c_0^2-r_0^2}\sqrt{H^2r_0^2
+\sin^2\theta}}\
\phi'^5 -\frac{4H^2r_0^2(H^2c_0^2+1)}{H^2r_0^2+\sin^2\theta}\
\phi^2\nonumber\\
&+&\frac{2H^2r_0^2\omega\cos\theta\dot\theta}{(H^2r_0^2+\sin^2\theta)^{3/2}}
\
\phi^4=0 \\
&&-\ddot\phi^3+{\phi''}^3-\frac{2c_0\omega}{\sqrt{c_0^2-r_0^2}}\
\dot\phi^1+\frac{2r_0\dot\theta
\sqrt{1+H^2r_0^2}}{\sqrt{c_0^2-r_0^2}\sqrt{H^2r_0^2+\sin^2\theta}}\
\dot\phi^5\nonumber\\
&-&\frac{2\sqrt{1+H^2r_0^2}\sin\theta}{\sqrt{H^2r_0^2+\sin^2\theta}}\
\phi'^4+
\frac{4H^2r_0^2(1+H^2r_0^2)}{H^2r_0^2+\sin^2\theta}\ \phi^3\nonumber\\
&+&\frac{2H^2r_0\cos\theta\sin\theta
\sqrt{1+H^2r_0^2}\sqrt{c_0^2-r_0^2}}{(H^2r_0^2+\sin^2\theta)^{3/2}}\
\phi^5=0\\
&&
-\ddot\phi^4+{\phi''}^4+\frac{2\omega\sin\theta}{\sqrt{H^2r_0^2+\sin^2\theta
}}\
\dot\phi^2
+\frac{2\sqrt{1+H^2r_0^2}\sin\theta}{\sqrt{H^2r_0^2+\sin^2\theta}}\
\phi'^3\nonumber\\
&-&\frac{2H^2r_0\cos\theta\sqrt{c_0^2-r_0^2}}{H^2r_0^2+\sin^2\theta}\
\phi'^5
+\frac{4H^2r_0^2\omega\dot\theta\cos\theta}{(H^2r_0^2+\sin^2\theta)^{3/2}}\
\phi^2\nonumber\\
&+&\frac{H^2r_0^2[-3H^2c_0^2+5H^2r_0^2+2H^4r_0^4-H^4c_0^2r_0^2]}
{(H^2r_0^2+\sin^2\theta)^2}\ \phi^4\nonumber\\
&+&\frac{H^2r_0^2[(2-2H^2r_0^2+2H^2c_0^2)\sin
^2\theta
-\sin^4\theta]}{(H^2r_0^2+\sin^2\theta)^2}\ \phi^4=0
\end{eqnarray}\begin{eqnarray}
&&-\ddot\phi^5+{\phi''}^5-\frac{2r_0\dot\theta\sqrt{1+H^2r_0^2}}{\sqrt{c_0^2
-r_0^2}\sqrt
{H^2r_0^2+\sin^2\theta}}
\ \dot\phi^3\nonumber\\
&-&\frac{2r_0\omega\dot\theta}{\sqrt{c_0^2-r_0^2}\sqrt{H^2r_0^2+\sin^2\theta
}}\
\phi'^2
+\frac{2H^2r_0\cos\theta\sqrt{c_0^2-r_0^2}}{H^2r_0^2+\sin^2\theta}\ \phi'^4
\nonumber\\
&+&\frac{4H^2r_0\sin\theta\cos\theta\sqrt{c_0^2-r_0^2}\sqrt{1+H^2r_0^2}}{(H^
2r_0^2+\sin^2\theta)^{3/2}}\phi^3\nonumber\\
&+&\frac{H^2(c_0^2-r_0^2)(-H^2r_0^2+2(H^2r_0^2+1)\sin^2\theta-\sin^4\theta)}
{(H^2r_0^2+\sin^2\theta)^2}\phi^5=0\label{4.17} \\
&&-\ddot\phi^6+{\phi''}^6+(2\sin^2\theta-H^2(c_0^2-2r_0^2))\phi^6=0\\
&&-\ddot\phi^7+{\phi''}^7+(2\sin^2\theta-H^2(c_0^2-2r_0^2))\phi^7=0\\
&&-\ddot\phi^8+{\phi''}^8+(2\sin^2\theta-H^2(c_0^2-2r_0^2))\phi^8=0\label{4.20}
\end{eqnarray}
To solve eqs.(\ref{4.13})-(\ref{4.20}), we make Fourier expansions
\begin{equation}\label{575}
\phi^i(\tau,\sigma)=\sum_{n}e^{in\sigma}\phi_n^i(\tau)
\end{equation}
where
${\phi_n^i}^{*}=\phi_{- n}^i$. We also insert the explicit solution
(\ref{2.9})
and define $\kappa=Hc_0$. The equations are now parametrised by the two
dimensionless
parameters $ (A, \kappa) $, where according to eq.(\ref{2.10}) $\kappa\geq
A$.

Notice that we have 5 coupled equations for $\phi_n^1 - \phi_n^5$, and 3
decoupled
equations for the others. Let us first see what we can say about the 3
identical
decoupled equations. They are given by (we skip the $i$ index)
\begin{equation}
\ddot\phi_n+\Bigg(A^2+
n^2-2A^2\mbox{sn}^2(\tau|A^2)
\Bigg)\phi_n=0 \ , \ \ A\leq 1
\end{equation}
\begin{equation}
\ddot\phi_n+\Bigg(A^2+
n^2-2\mbox{sn}^2(A\tau|1/A^2) \Bigg)\phi_n=0\ , \ \ A\geq 1
\end{equation}
For $n\neq 0 $,
the factor in front of $\phi_n$ is periodic and positive, making the solution 
stable.
The $n=0$ mode is just a zero-mode redefining the unperturbed solution.

Thus we are left with the first 5 equations.

\section{Analytical Results}
\setcounter{equation}{0}

First we look at the equations in some special limits. For $A=0$,
corresponding to no pulsation, we get

\begin{eqnarray}
&&\ddot\phi^1_n+n^2\phi^1_n-\sqrt{8(1+\kappa^2)}\dot\phi^3_n-\sqrt{4(2+\kappa^
2)} in\phi^2_n=0\\
&&\ddot\phi^2_n+n^2\phi^2_n+\sqrt{4(2+\kappa^2)}
in\phi^1_n+4(1+\kappa^2)\phi^2_n=0\\
&&\ddot\phi^3_n+n^2\phi^3_n+\sqrt{8(1+\kappa^2)}\dot\phi^1_n-2(2+\kappa^2)\phi
^3_n=0\\
&&\ddot\phi^4_n+n^2\phi^4_n+2 in\phi^5_n+\phi^4_n=0\\
&&\ddot\phi^5_n+n^2\phi^5_n-2 in\phi^4_n+\phi^5_n=0
\end{eqnarray}
The first 3 equations are of course the same as in \cite{tseytlin3},
and are stable for $\kappa^2\leq 5/2$. We are then left with the
last two equations which can be written
(we define $p_n^4=\dot{\phi_n^4}$ and $p_n^5=\dot{\phi_n^5}$)
\begin{eqnarray}
\left( \begin{array}{c}
\dot\phi_n^4\\
\dot p_n^4\\
\dot\phi_n^5\\
\dot p_n^5
\end{array} \right)
+
\left( \begin{array}{cccc}
0 & -1 & 0 & 0 \\
1+n^2 & 0 & 2in & 0 \\
0 & 0 & 0 & -1 \\
-2in & 0 & 1+n^2 & 0
\end{array} \right)
\left( \begin{array}{c}
\phi_n^4\\
p_n^4\\
\phi_n^5\\
p_n^5
\end{array} \right)
=
\left( \begin{array}{c}
0\\
0\\
0 \\
0
\end{array} \right)
\end{eqnarray}
To have stability we must have that all eigenvalues of the matrix are
imaginary.
The eigenvalues are
\begin{eqnarray}
\lambda_+=\pm i|1-n|\nonumber\\
\lambda_-=\pm i|1+n|
\end{eqnarray}
meaning stability. The modes corresponding to $n=\pm 1$ are
zero-modes. In conclusion,
for $A=0$ we have stability for $\kappa^2\leq 5/2$, corresponding to spin
(\ref{2.12})
\begin{eqnarray}
S\leq\frac{5\sqrt{7}}{8\sqrt{2}H^2\alpha'}
\end{eqnarray}
in agreement with \cite{tseytlin3}.

Another special case of importance is $A=1$, corresponding to a string
which starts at one of the poles and approaches the equator for
$\tau\rightarrow\infty$.
Unfortunately, the equations do not really simplify in this limit.
However, if we only look at the asymtotic equations, we get

\begin{eqnarray}
&&\ddot\phi^1_n+n^2\phi^1_n-2(\kappa
in\phi^2_n+\sqrt{2}\kappa\dot\phi^3_n)=0 \\
&&\ddot\phi^2_n+n^2\phi^2_n+2(\kappa
in\phi^1_n+\sqrt{2}\dot\phi^4_n)+4(\kappa^2-1)\phi^2_n=0\\
&&\ddot\phi^3_n+n^2\phi^3_n+2(in\phi^4_n+\sqrt{2}\kappa\dot\phi^1_n)-2(\kappa^
2-1)\phi^3_n=0\\
&&\ddot\phi^4_n+n^2\phi^4_n-2(in\phi^3_n+\sqrt{2}\dot\phi^2_n)=0\\
&&\ddot\phi^5_n+(n^2-1)\phi^5_n=0
\end{eqnarray}
It is of course somewhat dangerous to take $\tau\rightarrow\infty$ in the
equations,
since it corresponds to setting a
simple plane pendulum in the vertical upright position,
c.f.
the comments after eq.(\ref{2.10}). But we will see that some information
can be obtained anyway.

The last equation is solved by trigonometric functions, except for $n=0, \
\pm 1$.
We note that
$|n|=1$ are zero-modes, and $n=0$ just gives the expected exponential
divergence $\delta\theta \sim e^{\tau}$, which is an artifact of taking $
\theta=\pi/2$ in the equations,
c.f. the simple plane pendulum in the vertical upright position.

Let us now look at the first 4 equations and apply the method of the
previous case
\begin{eqnarray}
\left( \begin{array}{c}
\dot\phi_n^1\\
\dot p_n^1\\
\dot\phi_n^2\\
\dot p_n^2\\
\dot\phi_n^3\\
\dot p_n^3 \\
\dot\phi_n^4 \\
\dot p_n^4
\end{array} \right) +{\cal A}
\left( \begin{array}{c}
\phi_n^1\\
p_n^1\\
\phi_n^2\\
p_n^2\\
\phi_n^3\\
p_n^3 \\
\phi_n^4\\
p_n^4
\end{array} \right)
=
\left( \begin{array}{c}
0\\
0\\
0 \\
0\\
0\\
0 \\
0 \\
0
\end{array} \right)
\end{eqnarray}
where the matrix ${\cal A}$ is
\begin{equation}
\left( \begin{array}{cccccccc}
0 & -1 & 0 & 0 & 0 & 0 & 0 & 0\\
n^2 & 0 & -2\kappa in & 0 & 0 & -2^{3/2}\kappa & 0 & 0 \\
0 & 0 & 0 & -1 & 0 & 0 & 0 & 0 \\
2\kappa in & 0 & n^2+4(\kappa^2-1)& 0 & 0 & 0 & 0 & 2^{3/2}\\
0 & 0 & 0 & 0 & 0 & -1 & 0 & 0\\
0 & 2^{3/2}\kappa & 0 & 0 & n^2-2(\kappa^2-1) & 0 & 2in & 0\\
0 & 0 & 0 & 0 & 0 & 0 & 0 & -1 \\
0 & 0 & 0 & -2^{3/2} & -2in & 0 & n^2 & 0
\end{array} \right)
\end{equation}
Again we want to find the condition for which all the eigenvalues $\lambda$
are imaginary.
For $\lambda^2=l$ the characteristic polyomium is
\begin{eqnarray}
p(l)=&&l^4+l^3(4n^2+10\kappa^2+6)+l^2(6n^4+24\kappa^4+32\kappa^2+18\kappa^2n
^2+6n^2+8)\nonumber \\
&&+l(4n^6-6n^4+16n^2+6n^4\kappa^2+24\kappa^4n^2+24\kappa^2n^2)\nonumber\\
&& +n^8-6n^6+8n^4-2n^6\kappa^2+8n^4\kappa^2
\end{eqnarray}
We first show that the 3 extrema fall at negative $l$, i.e. that the roots
of
\begin{eqnarray}
p'(l)=&&4l^3+3l^2(4n^2+10\kappa^2+6)\nonumber \\
&&+2l(6n^4+24\kappa^4+32\kappa^2+18\kappa^2n^2+6n^2+8)\nonumber \\
&& +(4n^6-6n^4+16n^2+6n^4\kappa^2+24\kappa^4n^2+24\kappa^2n^2)
\end{eqnarray}
are negative. For that purpose, it is enough to show that the 2 extrema of
$p'(0)$ fall at negative $l$, and that
$p'(0)>0$.
We get
\begin{equation}
p''(l)=12l^2+6l(4n^2+10\kappa^2+6)+2(6n^4+24\kappa
^4+32\kappa^2+18\kappa^2n^2+6n^2+8)=0
\end{equation}
It is easily seen that both of the solutions of this equation fulfill the
requirement for arbitrary $n, \kappa$.
We then have to look at $p'(0)$
\begin{equation}
p'(0)=4n^6-6n^4+16n^2+6n^4\kappa^2+24\kappa^4n^2+24\kappa^2n^2
\end{equation}
Given that the last three terms are positive, and that the first three can
be written as
\begin{equation}
4n^6-6n^4+16n^2=n^2(2n^2-2)^2+2n^4+12n^2
\end{equation}
it is seen that $p'(0)>0$. Now we just have to find the condition that
$p(0)\geq 0$, which leads to
\begin{equation}
(4-n^2)(2+2\kappa^2-n^2)\geq 0
\end{equation}
For $n=0,1,2$, $\kappa$ is arbitrary.
For $n=3,4,5,...$ we get
$\kappa\leq\sqrt{\frac{7}{2}},\sqrt{7},\sqrt{\frac{23}{2}}$,...,
respectively. 
Thus we end up with $\kappa ^2\leq 7/2$, corresponding to the spin
\begin{equation}
S\leq \frac{15}{8\sqrt{2}H^2\alpha'}
\end{equation}
indicating, that we have better stability as compared to the $A=0$ case.

There is actually another case where we can solve the equations
analytically,
namely when $\kappa=A$. As seen from (\ref{2.10}) and (\ref{2.12}), this
limit corresponds to zero spin, $S=0$.
We get the equations for $i=1,2,3,4$

\begin{eqnarray}
&&\ddot \phi^1_n+n^2\phi^1_n - \sqrt{4(1+A^2)}\dot\phi^3_n -2in\phi^2_n=0\\
&&\ddot \phi^2_n+n^2\phi^2_n + \sqrt{4(1+A^2)}\dot\phi^4_n +2in\phi^1_n=0\\
&&\ddot \phi^3_n+n^2\phi^3_n + \sqrt{4(1+A^2)}\dot\phi^1_n +2in\phi^4_n=0\\
&&\ddot \phi^4_n+n^2\phi^4_n - \sqrt{4(1+A^2)}\dot\phi^2_n -2in\phi^3_n=0
\end{eqnarray}
And for $i=5$
\begin{eqnarray}
&&\ddot \phi^5_n+n^2\phi^5_n
-\frac{2-A^2\mbox{sn}^2(\tau|A^2))}{\mbox{sn}^2(\tau|A^2)}\phi^5_n=0\ , \ \
A\leq 1
\\
&&\ddot \phi^5_n+n^2\phi^5_n
-\frac{A^2(2-\mbox{sn}^2(A\tau|1/A^2))}{\mbox{sn}^2(A\tau|1/A^2)}\phi^5_n=0\
, \ \ A\geq 1
\end{eqnarray}
The first four equations
can be
shown to lead to oscillating solutions only, (and zero-modes) by the same
method as used before,
so we skip the details.

Since the term in front of $\phi_n^5$
is periodically minus infinity,
the equations for $\phi_n^5$ lead to instabilities.
It is not surprising that the pulsating strings become classically unstable
for $S\approx 0$.
Notice that the length of the unperturbed string is
\begin{eqnarray}
L(\tau)= \int_0^{2\pi}\frac{ds}{d\sigma}d\sigma=\frac{2\pi}{H}
\sqrt{H^2r_0^2+\sin^2\theta(\tau)}
\end{eqnarray}
$S\approx 0$ corresponds to $r_0 \approx 0$, which gives $L(\tau) = 2\pi
H^{-1} | \sin\theta(\tau)| $.
It follows that the string oscillates between its maximal size and zero
size.
It is well-known that the classical perturbations blow up when a circular
string collapses to
a point. But, of course, the classical approximation cannot be trusted in
this limit.

\section{Numerical Analysis} \setcounter{equation}{0}

Given the equations as they stand, eqs.(\ref{4.13})-(\ref{4.17}),
we have 5 coupled second order ordinary differential equations for the
Fourier components
$\phi^i_n$.

Needless to say, except for the few special cases considered in the
previous section,
any attempts at solving the equations analytically end
up being an exercise in futility. The correct way of going about the
problem is to solve the equations
numerically.
In what follows, we first convert the 5 second order ordinary differential
equations to 10 first order ordinary differential equations,
and then solve them numerically, using an appropriate algorithm.
A good safe choice is the classic fourth order Runge Kutta method for
step size $h$.

In order to run this algorithm, we must provide the step size $h$, and the
total number of steps, $N$.
In addition to that, we must also supply boundary conditions. The final
data was
analysed by printing the 10 graphs
for the 5 complex solutions, $\phi^1_n-\phi^5_n$.

All the aforementioned issues have been dealt with on a trial and error
basis. The step size has been made gradually
smaller, until there were no noticeable changes in the output, i.e., no
changes on the fifth decimal place.
In terms of $N$, we looked at the graphs and determined, that roughly
after 100000 points, it was clear whether
the functions diverged or not.

Ideally, the stability of the equations should be tested for an infinite
number of possible boundary conditions,
but since that is impossible, we have had to content ourselves with being
able to only check a limited number of
boundary conditions for the solutions.
We have not made a major issue out of it, and have only varied the
boundary conditions for a couple
of stable and unstable solutions.
This minor analysis showed, in terms of the behaviour of the
functions for large $\tau$, total insensitivity to varying the boundary
conditions.
Given that we are dealing with complex functions, we have chosen finite
different complex boundary conditions around unity
for all the functions.

We now turn to the numerical results. For each $(A, n) $ we integrate the
5 second order
equations
for $Hc_0 =\kappa> A$, to determine whether we have stability or not.
Typically we have stability up to a critical $\kappa$, beyond which we have
instability.
An example is shown in Figures 1 and 2, for $(A, n) =(2, 3)$.
In this case we have stability up to $\kappa\approx 2.54$,
and instability from $\kappa\approx 2.56$, which in terms of the spin is
\begin{equation}
S\leq 1.71
({H^2\alpha'})^{-1},\ \ \ A=2
\end{equation}
To find the critical $\kappa$ which separates
the stable and unstable solutions,
we now let $A$ and $n$ run. The corresponding critical spin $S$,
obtained from (\ref{2.12}), is plotted as a function of $A$ in Figure 3.

For $A=0,1$ we found, analytically,
that the modes $n=0, 1, 2$ are either zero-modes or stable. This turns out
to hold for arbitrary $A$.
We also saw in the same cases that
$n=3$ was the most unstable mode.
This also turns out to be the case for arbitrary $A$; see again Figure
3.

More precisely, in Figure 3, we show the maximal spin for which we have
stability as a function of $A$. We only show the curves for the modes
$n=3, 4, 5$,
just to show that $n=3$ is the most unstable mode, as already mentioned.

The numerical results are of course in agreement with the analytical
results
for $A=0, 1$, and from Figure 3 we can then finally conclude that
the inclusion of pulsation leads to better stability. In other words,
the more pulsation we have (the higher $A$), the higher spin we can allow
and still have stability.

\section{ Concluding Remarks}
\setcounter{equation}{0}
In conclusion, we have shown that when we include pulsation in $S^5$ 
\cite{patricio},
the resulting string solution has better stabillity properties than the 
original
two-spin string solution in $AdS_5$ \cite{tseytlin3}. It would be interesting 
to
see if pulsation could also improve the stabillity properties of the multi 
R-charge
solutions in $S^5$ \cite{tseytlin3}. Unfortunately, however, the equations for 
the perturbations
become even more complicated than our eqs.(\ref{4.13})-(\ref{4.20}), meaning 
that the analysis will demand extensive numerical work.
\vskip 24pt
{\bf Acknowledgements}:\\
We would like to thank N. Sibani for secretarial help, and L. Glazov for
help with the numerical work.
\newpage \appendix

\section{Appendix}
\setcounter{equation}{0}
The second
fundamental form and normal fundamental form
give (we only show the non-zero components)

\begin{eqnarray}
K^3_{\tau \tau}&=&K^3_{\sigma \sigma }=-r_0 \sqrt{1+H^2r_0^2} \\
K^5_{\tau \tau}&=&K^5_{\sigma \sigma }=\frac{-\sqrt{c_0^2-r_0^2}\
\sin\theta \cos\theta}{\sqrt{H^2r_0^2+ \sin^2\theta}} \\
K^2_{\tau \sigma }&=&\omega r_0 \\
K^4_{\tau \sigma }&=& \frac{-r_0 \cos \theta
\dot{\theta}}{\sqrt{H^2r_0^2+\sin^2\theta }}\\
\mu^{21}_\sigma &=&-\mu^{12}_\sigma
=\frac{-c_0\sqrt{1+H^2r_0^2}}{\sqrt{c_0^2-r_0^2}} \\
\mu^{25}_\sigma &=&-\mu^{52}_\sigma = \frac{r_0 \omega \dot{\theta}}
{\sqrt{c_0^2-r_0^2}\sqrt{H^2r_0^2+\sin^2\theta }}\\
\mu^{34}_\sigma &=&-\mu^{43}_\sigma =\frac{-\sqrt{1+H^2r_0^2}\
\sin\theta}{\sqrt{H^2r_0^2+\sin^2\theta}} \\
\mu^{45}_\sigma &=&-\mu^{54}_\sigma
=\frac{-H^2r_0\cos\theta\sqrt{c_0^2-r_0^2}}{H^2r_0^2+\sin^2\theta } \\
\mu^{31}_\tau &=&-\mu^{13}_\tau = \frac{c_0\omega}{\sqrt{c_0^2-r_0^2}}
\end{eqnarray}
\begin{eqnarray}
\mu^{42}_\tau &=&-\mu^{24}_\tau =
\frac{-\omega\sin\theta}{\sqrt{H^2r_0^2+\sin^2\theta }}\\
\mu^{53}_\tau &=&-\mu^{35}_\tau =\frac{r_0\dot{\theta} \sqrt{1+H^2r_0^2}}
{\sqrt{c_0^2-r_0^2}\sqrt{H^2r_0^2+\sin^2\theta}}
\end{eqnarray}
Now the Riemann component terms

\begin{eqnarray}
R_{\mu\rho\nu\sigma}N^\mu_i N^\nu_j(X'^\rho X'^\sigma-\dot{X}^\rho
\dot{X}^\sigma )
\end{eqnarray}
lead to
\begin{eqnarray}
R_{\mu\rho\nu\sigma}N^\mu_1 N^\nu_1(X'^\rho X'^\sigma-\dot{X}^\rho
\dot{X}^\sigma )&=&-H^2c_0^2\\
R_{\mu\rho\nu\sigma}N^\mu_2N^\nu_2(X'^\rho X'^\sigma-\dot{X}^\rho
\dot{X}^\sigma )&=&-H^2c_0^2\\
R_{\mu\rho\nu\sigma}N^\mu_3N^\nu_3(X'^\rho X'^\sigma-\dot{X}^\rho
\dot{X}^\sigma )&=&-H^2c_0^2\\
R_{\mu\rho\nu\sigma}N^\mu_4 N^\nu_4(X'^\rho X'^\sigma-\dot{X}^\rho
\dot{X}^\sigma )&=& H^2(2r_0^2-c_0^2) \\
R_{\mu\rho\nu\sigma}N^\mu_5N^\nu_5(X'^\rho X'^\sigma-\dot{X}^\rho
\dot{X}^\sigma )&=&-H^2(2r_0^2-c_0^2) \\
R_{\mu\rho\nu\sigma}N^\mu_6 N^\nu_6(X'^\rho X'^\sigma-\dot{X}^\rho
\dot{X}^\sigma )&=&2\sin^2\theta-H^2(c_0^2-2r_0^2)\\
R_{\mu\rho\nu\sigma}N^\mu_7 N^\nu_7(X'^\rho X'^\sigma-\dot{X}^\rho
\dot{X}^\sigma )&=&2\sin^2\theta-H^2(c_0^2-2r_0^2)\\
R_{\mu\rho\nu\sigma}N^\mu_8N^\nu_8(X'^\rho X'^\sigma-\dot{X}^\rho
\dot{X}^\sigma )&=&2\sin^2\theta-H^2(c_0^2-2r_0^2)
\end{eqnarray}

\newpage
\begin{center}
Figure Captions
\end{center}
\vskip 2 cm
\noindent
Figure 1. Example of a stable solution for $A=2, \ \kappa=2.54, \ n=3$. All
perturbations are finite oscillating functions.
\vskip 1 cm
\noindent
Figure 2.
Example of an unstable solution for $A=2, \ \kappa=2.56, \ n=3$. Some
perturbations blow up.
\vskip 1 cm
\noindent
Figure 3. Maximal spin S (in units of $(H^2\alpha')^{-1}$), for which we
have stable solutions, as a function of $A$.

\end{document}